**Diagnosing the magnetic field-tuned symmetry nature of the topological electrons by conductance fluctuations in bulk-insulating BiSbTeSe$_2$ devices**

Shuai Zhang[1], Faji Xie[1], Xing-Chen Pan[1], Yuyuan Qin[1], Lu Cao[1], Xuefeng Wang[2], Fengqi Song[1,*] and Baigeng Wang[1]

[1]National Laboratory of Solid State Microstructures, College of Physics and Collaborative Innovation Center of Advanced Microstructures, Nanjing University, Nanjing 210093, China

[2]National Laboratory of Solid State Microstructures, School of Electronic Science and Engineering and Collaborative Innovation Center of Advanced Microstructures, Nanjing University, Nanjing, 210093, China

[*] Corresponding author. Email: songfengqi@nju.edu.cn. Fax: +86-25-83595535

**Abstract**

We extract the quantum conductance fluctuations and study its magnetic field dependence in the gate-dependent transport of the topological electrons in bulk-insulating BiSbTeSe$_2$ devices. While increasing the magnetic field from 0 to 12 Tesla, the fluctuation magnitudes are found reduced by a ratio of $\sqrt{2}$ and form a quantized step. The step is observed both in n-type and p-type transport. It is also confirmed in the nonlocal measurements. This essentially demonstrates the breaking of the time reversal symmetry of the three-dimensional $Z_2$ topological insulators.

**Main text**

Statistic symmetry is one of the most elementary features of the electronic states and the symmetry operation determines the framework of condensed matter physics including band theory, phase transition theory and so on[1,2]. Even recently, in the topological classification of condensed matters, symmetry operation leads to versatile topological materials. Time reversal symmetry (TRS) protects the three-dimensional topological insulators (TIs)[3-5]. Space inversion symmetry protects the topological crystalline insulators[6]. Breaking either the space inversion symmetry or TRS leads to two different types of Weyl semimetals[7-10]. However, such an important feature of topological insulator has never been demonstrated and operated experimentally due to the substantial difficulty of closely tracking the electronic relativistic movement although intense efforts have been made and versatile devices have been demonstrated[11-16] since the discovery of TI. This becomes particularly urgent since the time-reversal symmetry is the crucial physics of three-dimensional $Z_2$ TIs.

Here we notice the magnitude of the quantum conductance fluctuation reveals the intrinsic symmetry[17] of the electronic states. It is universal conductance fluctuations (UCF) as manifested by the repeatable noise during the transport measurements[17-29]. It can be described by the random matrixes theory, where the electronic symmetry can be transferred during the matrixes operation and expressed in the root mean square (RMS) magnitudes of the UCF.

In this letter, we therefore fabricated the mesoscopic devices using the bulk-insulating topological insulators BiSbTeSe$_2$ (BSTS) and study the UCF dependence on the magnetic field. The UCF magnitudes are found reduced by a ratio of $\sqrt{2}$ and form a quantized step. This essentially indicates the breaking of the TRS in the three-dimensional $Z_2$ TIs.

The high-quality BiSbTeSe$_2$ crystals were grown by melting high-purity elements of bismuth, antimony, tellurium, selenium with a molar ratio of 1:1:1:2 at 850 °C for one day in evacuated quartz tubes, then cooling to room temperature over seven days. With mechanical exfoliation, the BSTS nanoribbons were transferred onto SiO$_2$/Si substrates. We fabricated the BSTS field-effect transistor devices by photolithography and electron-beam evaporation, followed by standard lift-off technique. The atomic force microscope (AFM) picture of the sample was shown in figure 1(a), with the white scale bar being 2 μm. The line marked with blue dash circle was plotted to show the height of the BSTS sample. The height is 92 nm, and width ($W$) is 0.9 μm. The distance between the two nearest electrodes ($L$) is 1.8 μm. The parameters of the two devices are shown in Table I. The transport of the BSTS samples is dominated by the surface state. We successfully observed quantum Hall effect in the samples[30]. And the gate-tuned bipolar feature can be used to design frequency multiplier, which we have also achieved. The typical temperature ($T$)-dependent resistance curve is shown in figure 1(c). At high $T$, it exhibits the insulating behavior. And when $T < 60$K, the metallic behavior means that topological surface state (TSS) is dominate.

As an important mesoscopic quantum interference phenomenon[31], The UCF is a kind of irregular fluctuations in the conductance, which is repeatable. It comes from the interference of Feynman path of electron diffusion. When the device size is within the coherence length, the RMS of UCF magnitude[19] ($\delta G_{RMS}$) would be in the order of $e^2/h$, where $e$ is the electronic charge and $h$ is the Planck constant. This is a universal constant, independent of the sample material, size and the disorder degree. In figure 1(c), the magnetoconductance (MC) of the sample was shown at 1.8 K. Besides the weak antilocalization (WAL) effect, the conductance fluctuation (CF) is clearly exhibited. After subtracting the smooth background (red line), we get the CF pattern. By changing the temperature, we get the repeatable patterns, shown in figure 1(d). Also, the CFs under both positive and negative direction of magnetic fields has similar features. According to Hikami-Larkin-Nagaoka (HLN) theory[32], we can fit the WAL of the MC. The phase coherence length $l_\phi$ is about 122 nm at $T = 1.8$ K from the fitting. The coefficient of the channel $\alpha$ at this condition is about 1.1, indicating the independent top and bottom TSS transport channels.

The UCF features can also be observed in the nonlocal measurement using the configuration as shown in the insert of figure 1(e). Figure 1(e) exhibits the nonlocal MC and CF pattern at 1.8K. For clarity, the field dependent UCFs of nonlocal and local measurement are plotted in figure 1(f). We can see that the nonlocal CF patterns have the similar features at both positive and negative magnetic field, while they are different from the local CF pattern. Because the nonlocal CF is from the probe 1 and 2, while the local CF is from probe 2 and 3. The different parts of the sample have various impurity configurations. That is why the CF patterns are different.

The extracted signals can be understood by the autocorrelation function $F(\Delta B) =$

⟨δ$G(B)$δ$G(B+\Delta B)$⟩, where ⟨⋯⟩ represents the ensemble average. We can define δ$G_{RMS}$ = $[F(0)]^{1/2}$. We choose the CF with the magnetic field over 1 T to calculate the δ$G_{RMS}$ = 0.0086 $e^2/h$ at $T$ = 1.8 K. However, from the UCF theory of the TSS developed recently, the δ$G_{RMS}$ of the TSS should be in the range[23, 33] from 0.43 to 0.54 $e^2/h$. This is due to the size of the sample is much larger than the phase coherence length ($L > l_\phi$). The self-averaging modifies[25] the δ$G_{RMS}$ as δ$G_{RMS} \simeq \delta G_{RMS}^{in} \cdot \gamma \frac{l_\phi W^{1/2}}{L^{3/2}}$, where $\gamma$ is a suppression factor related to the symmetry of the system and is $1/2\sqrt{2}$ here, and δ$G_{RMS}^{int}$ is the intrinsic UCF amplitudes[34]. From this formula, we can get δ$G_{RMS}^{int} \approx$ 0.52 $e^2/h$. This evidences the observation of the UCF of TI.

The random matrixes theory describes the δ$G_{RMS}$ of an electronic state in an analytic form as follows[17]

$$\delta G_{RMS} \simeq c_d \frac{e^2}{h} \sqrt{ks^2/\beta} \tag{1}$$

where $c_d$ is a coefficient related to the dimension $d$, $k$ is the number of independent eigenvalue sequences, $s$ is the degeneracy of eigenvalue sequence and $\beta$ is a constant related to the statistical symmetry. Dyson introduced the Gaussian distribution of Hamiltonian and classified random with different symmetry into three ensembles which also called three universal symmetry classes[35]. Gaussian orthogonal ensemble (GOE) describes the system with both TRS and spin rotational symmetry. The Wigner-Dyson parameter $\beta = 1$ for this situation. Gaussian unitary ensemble (GUE) describes the TRS-broken system, and $\beta = 2$. Gaussian symplectic ensemble (GSE) describes the TRS system without spin rotational symmetry, i.e. there is the existence of spin-orbit coupling (SOC), and $\beta = 4$. For disordered metal system without SOC

(or with very weak SOC), it belongs to GOE in the absence of magnetic field, and the symmetry parameters are given by $s = 2$, $k = 1$ and $\beta = 1$. When the magnetic field exceeds the threshold of TRS, the system changes to GUE, and they become $s = 2$, $k = 1$ and $\beta = 2$. The TRS protects the topological number of $Z_2$ and is therefore the crucial feature of the three-dimensional TIs. Upon the strong magnetic field, the TRS will be broken. As a consequence[21], with the increase of magnetic field, UCF amplitude will be reduced by $\sqrt{2}$ times when TRS is broke ($B > B_C$).

The expected UCF step is observed when we study the UCF dependence on the magnetic field. The backgate voltage ($V_g$) dependent CF, i.e. the $\delta G$-$V_g$ here shown in figure 2. As we all known, the gate voltage can be used to modulate the carrier concentration in the field-effect transistor device, and therefore tune the position of the Fermi level, i.e., the coherence paths. So it is the Fermi level dependent UCF. Although the gate voltage dependent UCF was observed in the study of AB oscillations in TI[12], the systematic study has not been carried out. Here, we measured the $V_g$ dependent resistance from 0 to 12 T at $T = 1.8$ K, plotted in figure 2(a). In our system, $V_g$ only affects the bottom surface state due to the thick bulk. Therefore, a single TSS (bottom surface) is the host of this behavior. Five CF patterns (figure 2(b)) exhibit the similar detail features in a large range of magnetic field. The $V_g$-dependent nonlocal CFs are shown in figure 2(d), which is extracted from the resistance in figure 2(c). Similar to the field dependent CFs, the nonlocal CFs also have different features with the local CFs. Figure 2(e) shows the CF mapping with $V_g$ at various magnetic fields. In figure 2(f), as we proposed before[27], we observe the $\sqrt{2}$ times reduction of the $\delta G_{RMS}$ with the magnetic field increasing. At $B = 0$ T, the system belongs to GSE, so $\beta = 4$. With the $s$ being 2 and $k$ being 1, $\sqrt{ks^2/\beta}$ is then 1. With the magnetic field increasing, the TRS is broken, so there is a transition from GSE to GUE. And $s$ changes to 1, while $k$ keeps unchanged. At this situation, the value of $\sqrt{ks^2/\beta}$ becomes $1/\sqrt{2}$. The transition magnetic field ($B_C$) reflected the TRS broken. We can see that $B_C$ here is around 0.1 T. The magnetic length[20] $l_C$ of $B_C$ is $\sqrt{h/eB_C}$, which value is about 200 nm. Compared with the phase coherence length $l_\phi$ from WAL effect, the $l_C$ is consistent with $l_\phi$. We can see there is no another $\sqrt{2}$ times reduction with the magnetic field increasing like normal metal. This is the manifestation of the lack of Zeeman splitting due to the lack of spin degeneracy, although there is Zeeman shift[36-38] of the zero mode Landau level in TIs. The CF mapping also reflects the change in symmetry. When the magnetic field is lower than $B_c$, the peaks (red strips) and valleys (blue strips) of the CFs are in the same position, as shown in figure 2(c), especially the lower-right part of the figure. Once the TRS is broken, the positions are changed obviously.

To explore whether the carrier type would effect this law, we show another sample (labeled with 'B'), which exhibits the bipolar curves, and the Dirac point (DP) is near $V_g = -14$ V. Here, we show the resistance mapping with magnetic field and $V_g$ in figure 3(a). For the $V_g$-tuned conductance at $B = 0$ T, we can get the CF pattern (figure 3(b)). As show in figure 3(c, d), the $\sqrt{2}$ time reduction is also observed for the $\delta G_{RMS}$ evolution in both electron (the left side of DP) and hole regions (the right side of DP). One important thing is that the amplitude of the UCF in the electron region is larger than that in the hole region. By calculation, we get the $\delta G_{RMS}$ in the hole region (-60 ~ -20 V) is 0.012 $e^2/h$ at zero field, and in the electron region (-10~30V) it is 0.027 $e^2/h$. While in the two-dimensional TSS, the theory[31] gives UCF amplitude as $\delta G_{RMS} \approx l_\phi^{(4-d)/2}$ when $L \gg l_\phi$, where $d$ is the dimension of the system. We measured the magnetoresistance with different $V_g$ at $T = 1.8$ K. They are shown in figure 3(e). From the HLN fitting, we extract the coherence length $l_\phi$ in figure 3(f). It clearly exhibits that the $l_\phi$ in the hole region is smaller than in the electron region. The similar phase coherence length $l_\phi$ variation with $V_g$ is also observed[39] by other group. In graphene, the electron-hole asymmetry can be induced by strain and charged defects[40]. Therefore, the charged defects in the topological insulator can also induce the electron-hole asymmetry of the TSS. As a consequence, the $\delta G_{RMS}$ in the electron is larger than in the hole region. Also the mobility in the two regions is different. The field-effect mobility formula[41] is $\mu = \frac{L}{W}\frac{dG}{dV_g}\frac{1}{C_i}$, where $L$ is the length, $W$ is the width of the channel and $C_i$ is the capacitance of the dielectric layer. Here we take the value of $C_i$ as $11.5 \times 10^{-9}$ F/cm$^2$. From the $G$-$V_g$ data of 1.8 K, we extract the hole mobility $\mu_h = 380$ cm$^2$/Vs and the electron mobility $\mu_e = 740$ cm$^2$/Vs. We are therefore convinced the UCF $\sqrt{2}$ step is universal. This indicates the breaking of the TRS of a $Z_2$ TI.

In summary, we observe the UCFs with magnetic field and $V_g$ in both local and nonlocal configuration. Besides, the Fermi level dependent UCF is studied here. We observe a $\sqrt{2}$ times reduction in $\delta G_{RMS}$ with magnetic field increasing, which can be explained by the theory of the random matrix. This reveals the sympletic transport of topological electron states.

**Acknowledgements**

This work was supported by the National Key R&D Program of China (Grant No. 2017YFA0303200), the National Natural Science Foundation of China (Grants No. U1732273, U1732159, 91421109, 91622115, 11522432 and 11574217), the Natural Science Foundation of Jiangsu Province (Grants No. BK20160659), and the Opening Project of Wuhan National Pulsed Magnetic Field Center. We would also like to acknowledge the helpful assistance of the Nanofabrication and Characterization Center at the Physics College of Nanjing University, and thank Prof. Xin-Cheng Xie in Peking University and Prof. Hua Jiang in Soochow University for stimulating discussions.

**Figure Captions**

**Figure 1. Observation of the UCF in the magnetoconductance of BSTS devices.** (**a**) The AFM picture of the BSTS nanoribbon device. The white scale bar is 2 μm. The height of the line marked with a blue dash circle in the AFM picture. (**b**) The typical temperature-dependent resistance curve. (**c**) The magnetoconductance is measured at 1.8 K. After subtracting the smooth background (red line), we get the conductance fluctuation (CF). (**d**) The CF patterns are shown at 1.8K (both of the positive and negative direction of the magnetic field), 2K and 3K. They show the similar features clearly. They are offset by 0.06 $e^2/h$. (**e**) The nonlocal measurement of magnetoconductance and the CF pattern at 1.8K. The inset is the sketch of the nonlocal measurement. (**f**) The field dependent CFs of nonlocal and local measurement. Offset by 0.06 $e^2/h$.

**Figure 2. Observation of the magnetic-field driven $\sqrt{2}$ UCF step in the $V_g$-$G$ curves in the BSTS devices.** (**a**) The backgate voltage dependent resistance shows at 1.8 K with the magnetic field varying from 0 to 12 T. (**b**) Choosing the different magnetic field to plotting the CF patterns, offset by 0.02 $e^2/h$. (**c**) The gate voltage dependent resistant of nonlocal and local measurement. (**d**) The backgate-dependent CFs of nonlocal and local measurement. Offset by 0.02 $e^2/h$. (**e**) The CF mapping is from fig.2 (a) by subtracting the smooth background. (**f**) The renormalized RMS values of the CF amplitude vary with the magnetic field. They show $\sqrt{2}$ times decay with the magnetic field increasing.

**Figure 3. Confirming the $\sqrt{2}$ step of the UCF while changing the carrier types.**
**(a)** The resistance mapping with backgate voltage and magnetic field at 1.8 K for sample B. **(b)** The backgate-tuned conductance and CF pattern at $B = 0$ T. **(c)** The RMS of the UCF amplitude also obeys the $\sqrt{2}$ times reduction in hole region. The difference is that the amplitude value of UCF in the hole region is samller than in the electron region. **(d)** The RMS of the UCF amplitude also obeys the √2 times reduction in electron region. The difference is that the amplitude value of UCF in the hole region is smaller than in the electron region. **(e)** The magnetoresistance at different backgate voltage, $T = 1.8$ K. **(f)** The coherence length varying with backgate voltage.

**Table 1.** The parameters of the two samples.

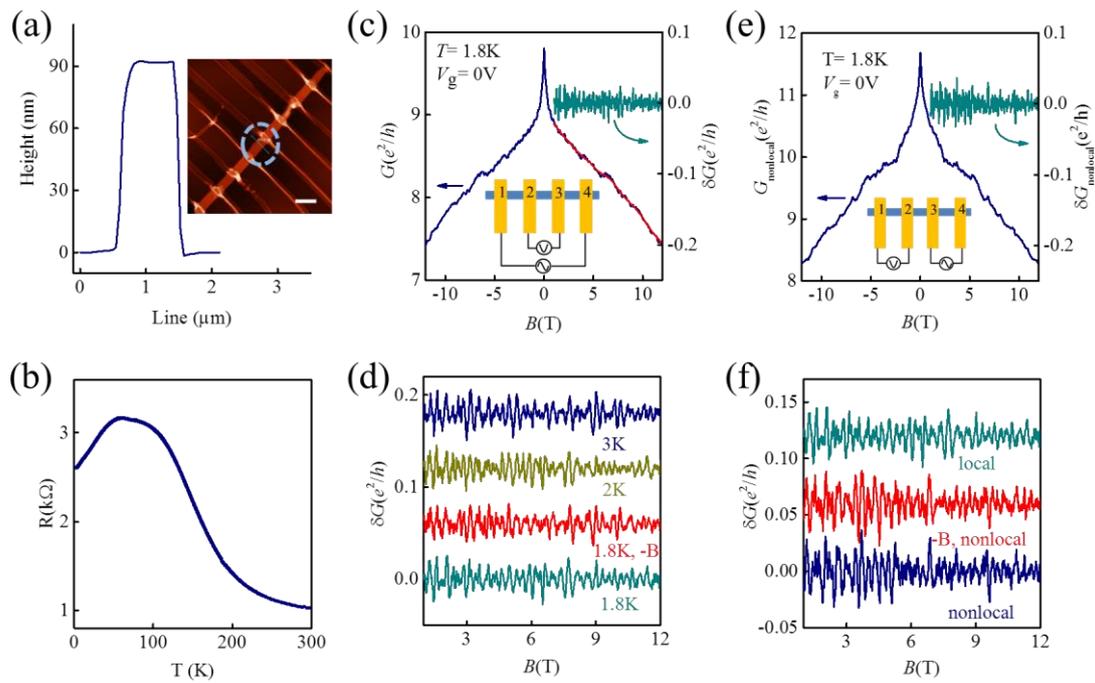

**S. Zhang et al, Figure 1**

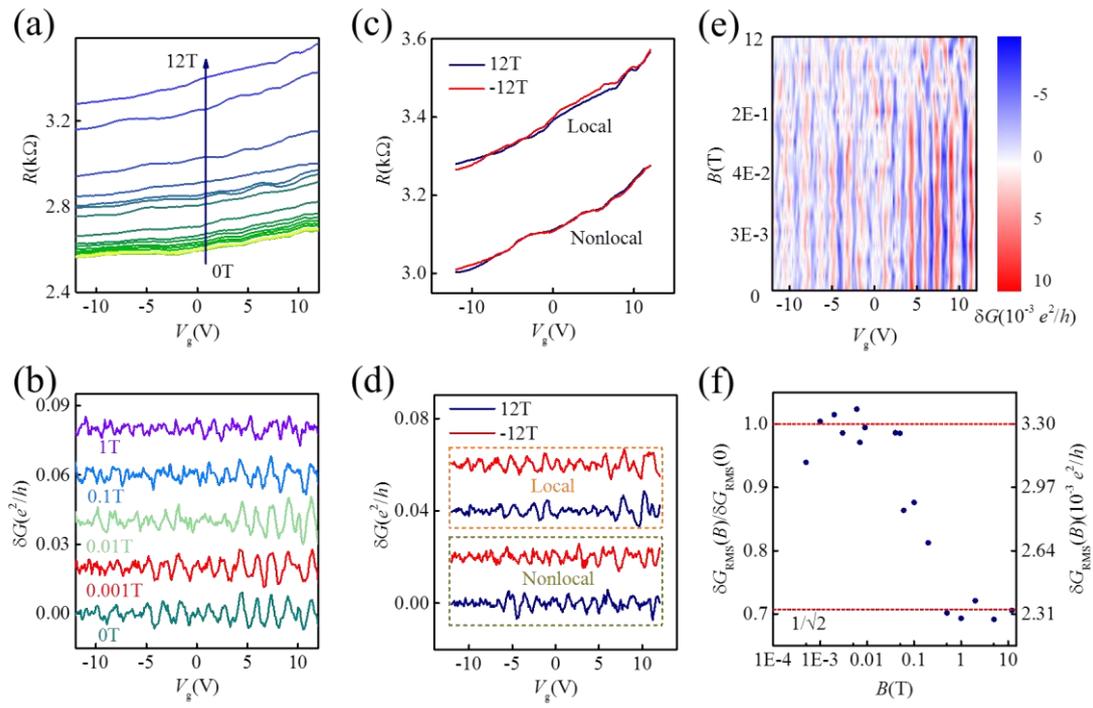

**S. Zhang et al Figure 2**

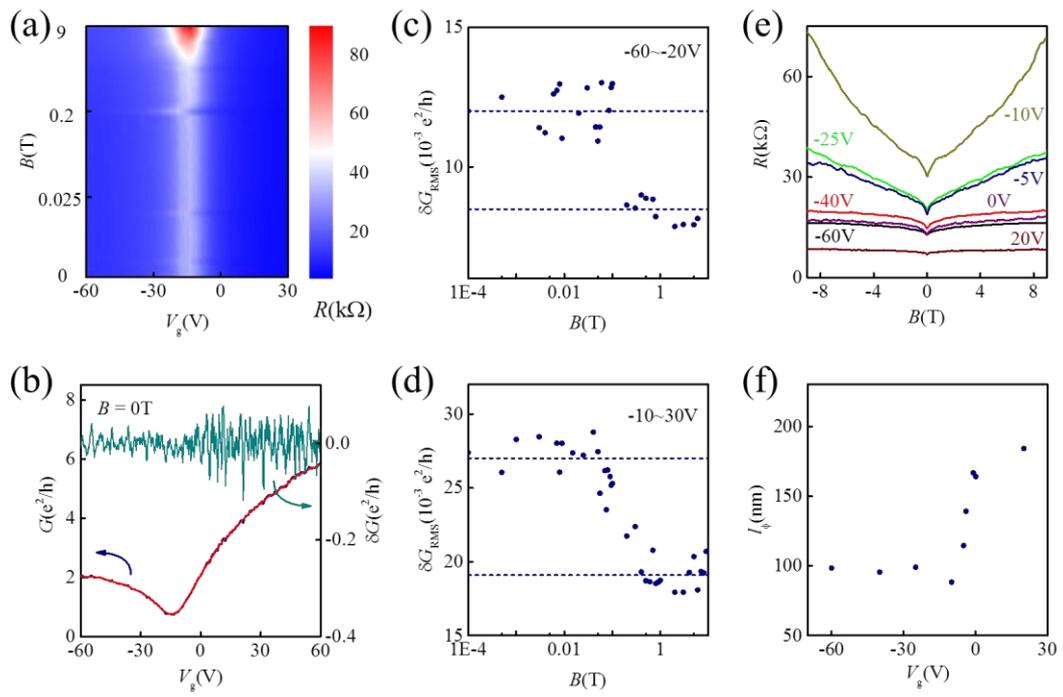

**S. Zhang et al Figure 3**

| Sample | A | B |
|---|---|---|
| Size ($L \times W$) | 1.8μm×0.9μm | 2.0μm×1.2μm |
| $R$ (2K) | 2.6 kΩ | 2.13 kΩ |
| $R$ (280K) | 1.05 kΩ | 2.73 kΩ |
| $\delta G_{RMS}$ ($B$=0) | 0.0033 $e^2/h$ (h) | 0.012 $e^2/h$ (h)<br>0.027 $e^2/h$ (e) |
| Field-effect mobility | 233 cm$^2$/Vs (h) | 380 cm$^2$/Vs (h)<br>740 cm$^2$/Vs (e) |
| Carrier type ($V_g$=0) | p | n |
| √2-Reduction | Yes | Yes |
| $B_C$ | ~0.1 T | ~0.1 T |

**S. Zhang et al Table 1**